\documentclass[showpacs,twocolumn,aps,preprintnumbers,letterpaper]{revtex4}
\usepackage{amsmath,amssymb}
\usepackage{epsfig}
\usepackage{graphicx}
\usepackage{amsmath}
\usepackage{amsfonts}
\usepackage{epstopdf}

\newcommand{\be}{\begin{equation}}
\newcommand{\ee}{\end{equation}}
\newcommand{\bear}{\begin{eqnarray}}
\newcommand{\eear}{\end{eqnarray}}
\newcommand{\ba}{\begin{array}}
\newcommand{\ea}{\end{array}}

\def\be{\begin{eqnarray}}
\def\ee{\end{eqnarray}}

\def\roughly#1{\mathrel{\raise.3ex\hbox{$#1$\kern-.75em%
\lower1ex\hbox{$\sim$}}}}

\def\la{{\Big<}}
\def\ra{{\Big>}}

\begin{document}

\title{Holographic Pomeron: Saturation and DIS}

\author{Alexander Stoffers}
\email{alexander.stoffers@stonybrook.edu}
\affiliation{Department of Physics and Astronomy, Stony Brook University, Stony Brook, New York 11794-3800, USA}

\author{Ismail Zahed}
\email{zahed@tonic.physics.sunysb.edu}
\affiliation{Department of Physics and Astronomy, Stony Brook University, Stony Brook, New York 11794-3800, USA}

\date{\today}

\begin{abstract}
We briefly review the approach to dipole-dipole scattering in holographic QCD developed in \cite{Basar:2012jb}
and based on a stringy Schwinger mechanism.  The Pomeron emerges through the exchange of closed 
strings between the dipoles and yields Regge behavior for the elastic amplitude.
We calculate curvature corrections to this amplitude in both a conformal and confining background, identifying the holographic direction with the virtuality of the dipoles. The {\it wee-dipole} density is related to the string tachyon diffusion in both virtuality and the transverse directions. We give an explicit derivation of the dipole saturation momentum both in the conformal and confining metric.
Our holographic result for the dipole-dipole cross section and the {\it wee-dipole} density in the conformal limit are shown to be identical in form  to the BFKL pomeron result when the non-critical string transverse dimension is $D_\perp=3$. The total dipole-dipole cross section is compared to DIS data from HERA.

\end{abstract}

\maketitle

\setcounter{footnote}{0}


\section{Introduction}

In perturbative QCD, dipole-dipole as onium-onium scattering has long been used to describe high energy scattering, \cite{Mueller:1989st, Mueller:1994gb,Mueller:1993rr,Mueller:1994jq,Iancu:2003uh, Nikolaev:1990ja,Nikolaev:1991et,Salam:1995zd,Salam:1995uy,Navelet:1996jx,Navelet:1997tx,GolecBiernat:1998js}. In the 1-pomeron exchange, this is equivalent to the BFKL approach~\cite{BFKL}. The scattering amplitude can be defined though the convolution of densities of the {\it wee-dipoles} originating from the parent dipoles and
diffusing along the rapidity direction  in transverse space. This fundamental diffusion was forseen long ago by Gribov and will
be referred to as Gribov's diffusion~\cite{GRIBOV}.

Holographic QCD offers a non-perturbative framework for discussing 
diffractive scattering at large number of colors $N_c$ and  strong t'Hooft 
coupling $\lambda=g^2N_c$.  A number of authors~\cite{Rho:1999jm,Basar:2012jb,Janik:1999zk,Janik:2001sc,Janik:2000aj, Janik:2000pp,Polchinski:2001tt, Polchinski:2002jw,Brower:2006ea,Brower:2007xg} 
have suggested that small angle hadron-hadron scattering in holographic
QCD can be understood as string exchange.  At large rapidity $\chi$, the elastic
scattering amplitude is purely imaginary.

Diffractive dipole-dipole scattering in holographic QCD is dominated by
closed-string pair creation at large $\chi$ and impact parameter 
${\bf b}_\perp$ \cite{Basar:2012jb}. This stringy Schwinger mechanism is
due to the occurence of a longitudinal "electric field"
$E/\sigma_T={\rm tanh}(\chi/2)$ triggered by the rapidity $\chi$ of
the scattering dipoles, with $\sigma_T$ the (fundamental) string tension.
The pair creation process is $e^{-1/\chi}$ suppressed 
as $\chi\rightarrow 0$, showing the discontinuous nature of the 
time-like versus space-like  surface exchange. The process
exhibits Gribov's diffusion in the transverse space owing to the emergence of a large
Unruh temperature and the subsequent dimensional reduction.

In curved AdS, the
scattering amplitude in holography is closely related to Gribov diffusion in curved space~\cite{Brower:2006ea,Brower:2007xg,
Cornalba:2009ax}. 
As the analysis of the Schwinger mechanism in~\cite{Basar:2012jb} was carried using string exchange in Witten's confining
background in the near horizon limit (flat space), it is important
to extend it to curved AdS space.  Below, we show that the extension to conformal AdS$_3$ 
(short for transverse AdS$_5$) yields a result that is similar to the one for the onium-onium scattering
amplitude following from the BFKL pomeron exchange in QCD \cite{Mueller:1989st, Mueller:1994gb,Mueller:1993rr,Mueller:1994jq,Iancu:2003uh, Nikolaev:1990ja,Nikolaev:1991et,Salam:1995zd,Salam:1995uy,Navelet:1996jx,Navelet:1997tx,GolecBiernat:1998js}.  The
differences being the pomeron intercept and diffusion constant.
We also discuss the concept of dipole saturation for conformal and confining  AdS$_3$ 
a point of intense interest both at HERA and present and future colliders. Recently,
the concept of saturation in holography was addressed  in~\cite{Hatta:2007cs, Hatta:2007he,
Cornalba:2008sp,Albacete:2008ze, Albacete:2008vv,Cornalba:2010vk} using different arguments.

In~ \cite{Basar:2012jb} it was suggested that in the regime $\chi > \lambda$ the string is far from critical and supersymmetric.
In this regime, the string worldsheet was argued to be bosonic with $D_\perp<8$, 8 being the
critical value for superstrings. Below we suggest $D_\perp=3$ based on comparison with the QCD BFKL pomeron, with 2 transverse spatial dimensions and 1 conformal dimension~\cite{Mueller:1989st, Mueller:1994gb,Mueller:1993rr,Mueller:1994jq,Iancu:2003uh, Nikolaev:1990ja,Nikolaev:1991et,Salam:1995zd,Salam:1995uy,Navelet:1996jx,Navelet:1997tx,GolecBiernat:1998js}.
In other words, the holographic pomeron is far from the graviton exchange suggested by the critical string analysis in~\cite{Polchinski:2001tt, Polchinski:2002jw,Brower:2006ea,Brower:2007xg}. The holographic pomeron in $D_\perp=3$ follows from an effective string theory in $D=3+2$ perhaps of the type advocated by Luscher~\cite{Luscher:1980ac}. While we will use holographic QCD as a model
throughout, a more systematic approach within holography and following Luscher's long string arguments maybe sought 
in AdS along the arguments in~\cite{Aharony:2010cx}.

In section 2, we briefly review the holographic Schwinger mechanism
for dipole-dipole scattering. In section 3, we argue that the induced Unruh temperature 
$T_{U}\approx \chi/2\pi {\bf b}_\perp$ on the string world-sheet causes the dual-open string to reduce dimensionally
from $D$ to $D_\perp$ dimensions. For large impact parameters ${\bf b}_\perp$, this
reduction is at the origin of the Gribov diffusion in $D_\perp$ dimensions.
In section 4, we suggest a holographic correspondence between dipole-dipole
scattering of varying sizes and parton-parton scattering of varying scales, whereby
the scale is identified with the holographic direction $z$~\cite{Polchinski:2001tt}. 
In section 5, we derive the {\it wee-dipole} density and define the saturation momentum for dipole-dipole scattering through the total inclusive cross section both for the conformal as well as the  confining geometry (conformal plus a wall).
For the conformal background, our results are compared to the perturbative QCD dipole-dipole scattering result using BFKL methods. We compare our results for the proton structure function $F_2$ to HERA data in section 6. Our conclusions are in section 7.

\section{Onium-Onium Scattering} 

QCD dipole-dipole scattering at large $\chi$ and weak coupling has been extensively discussed  by Mueller and others
\cite{Mueller:1989st, Mueller:1994gb,Mueller:1993rr,Mueller:1994jq,Mueller:1994gb,Iancu:2003uh, Nikolaev:1990ja,Nikolaev:1991et,Salam:1995zd,Salam:1995uy,Navelet:1996jx,Navelet:1997tx}. This approach to high-energy hadron-hadron
scattering was pioneered by Gribov~\cite{GRIBOV}. Typically, the scattering is viewed as a parent dipole of size $a$
depleting into a cascade of daughter ({\it wee-dipoles})  and smashing against a similar parent dipole of size $a'$ for
fixed income parameter ${\bf b}_\perp$.  The onium-onium
cross section in the single BFKL exchange reads~\cite{Mueller:1994gb}

\be
\sigma_{\rm tot}^{{\bf BKFL}} (\chi)=2\pi \frac{\lambda^{3/2}}{N_c^2}\,aa'\,\frac{e^{(\alpha^{\scriptscriptstyle \bf BFKL}-1)\chi}}{\left(4\pi {\bf D}^{\scriptscriptstyle \bf BFKL} \ \chi \right)^{1/2}} \ ,
\label{DDQCD}
\ee
with 
\be
\alpha^{\scriptscriptstyle \bf BFKL}=&&1+\frac{\lambda}{\pi^2}\,{\rm ln}\,2 \label{onium}\nonumber\\
{\bf D}^{\scriptscriptstyle \bf BFKL}=&&7 \lambda \zeta(3) /(8\pi^2)
\ee
the BFKL intercept and diffusion constant respectively. $\zeta$ is the Riemann zeta function. 

The holographic dipole-dipole scattering amplitude in a confining (Witten)  AdS background in the single funnel-exchange approximation was worked out in~ \cite{Basar:2012jb} in the near horizon limit or long strings.  In this work, the confining background
will be simplified to AdS with a wall with identical results for long strings. The dipole-dipole cross section for a single string exchange in the near horizon limit is~\cite{Basar:2012jb}

\be
\sigma_{\rm tot}(\chi)&=&\frac{g_s^2}{2 \alpha'}  \left(2 \pi^2 \alpha' \right)^{D_\perp /2}\,aa'\,\frac{e^{(\alpha_{\bf P}-1)\chi}}{(4 \pi {\bf D} \chi)^{D_\perp/2-1/2}} \ .
\label{HOLOX}
\ee
In leading order in $1/\sqrt{\lambda}$

\be
\alpha_{\bf P}=&&1+\frac{D_\perp}{12} \nonumber\\
{\bf D}=&& \frac{\alpha'}2=\frac 1{2\sqrt{\lambda}}
\label{HOLO2}
\ee
are the holographic intercept and diffusion constant respectively. For AdS with a wall, the {\it effective} string tension will be defined as

\be
g_s\equiv \kappa \frac{1}{{4\pi\alpha'}^2 N_c}\equiv\kappa \frac{{\lambda}}{4\pi N_c} \ ,
\label{GS}
\ee
with $\alpha'=1/\sqrt{\lambda}$ the string tension (in units of the AdS radius). We note that the $1/N_c$ dependence
of the string coupling does not depend on the nature of the holographic model, but the $\lambda$ dependence does.
The parameter $\kappa$ follows from the unfixed overall constant in the derivation of the dipole-dipole amplitude,
see eqn. (32) in~\cite{Basar:2012jb}, where it was absorbed in the definition of the dipole size.
Here we have chosen to absorb it into the definition of the string
coupling instead, hence the label {\it effective}. Below, it will be fixed phenomenologically to $\kappa=2.5$.

A comparison of (\ref{HOLO2}) to (\ref{DDQCD}) suggests that effectively $D_\perp=3$, an observation that
will be iterated below. In a way, the results of ~\cite{Basar:2012jb} can be thought as following 
from a non-critical string theory in $D_\perp+2$ dimensions,  as perhaps advocated by Luscher a while ago \cite{Luscher:1980ac}.  
In this effective string theory, the Polyakov action is the leading and universal contribution for long strings. 

The dependence on $aa'$ in (\ref{DDQCD}) and (\ref{HOLOX}) is noteworthy as it differs from $a^2{a'}^2$ in perturbation
theory.  In the string description this follows readily from the fact that the string attachment to the dipole only depends on
$a$ and $a'$ at each end respectively. This is not the case for 2 gluons exchange in perturbation theory where each 
gluon attaches to $a$ and $a'$ respectively, thus $a^2$ and ${a'}^2$. In both cases the dipole-dipole cross section
vanishes as $a,a'\rightarrow 0$ as it should.

\section{Stringy Schwinger mechanism}

The eikonalized dipole-dipole scattering amplitude ${\cal T}$ in Euclidean
space takes the following form~\cite{NACHTMANN}
\be
\frac 1{-2is} {\cal T} (\theta , q ) \approx \,\int d {\bf b}_\perp \,\, e^{iq_{\perp}\cdot {\bf b}_\perp}\,{\bf WW} \ ,
\label{4X}
\ee
with
\be
{\bf WW}=\la \left({\bf W} (\theta, {\bf b}_\perp) -{\bf 1}\right) 
\left( {\bf W} (0,0) -{\bf 1}\right)\ra\,,
\label{CORR}
\ee
and
\be
{\bf W} (\theta, {\bf b}_\perp)= \frac 1{N_c} {\rm Tr} \left( {\bf P}_c {\rm exp}\left(ig\int_{{\cal C}_{\theta}}\,d\tau\,
A(x)\cdot v\right)\right)\, 
\label{5X}
\ee
is the normalized Wilson loop for a dipole, $\langle {\bf W}\rangle\equiv{\bf 1}$. In Euclidean geometry
${\cal C}_{\theta}$  is a closed rectangular loop of width $a$ that is
slopped at an angle $\theta$ with respect to the vertical imaginary time direction.
The two dimensional integral in (\ref{4X}) is over the impact parameter ${\bf b}_\perp$
with $t=-q_{\perp}^2$, and  the averaging is over the gauge configurations
using the QCD action.

Following ~\cite{Rho:1999jm,Basar:2012jb,Janik:1999zk,Janik:2001sc,Janik:2000aj, Janik:2000pp,Polchinski:2001tt, Polchinski:2002jw,Brower:2006ea,Brower:2007xg}, the averaging in (\ref{CORR}) will be sought in the context of holography.
For small size dipoles, the chief contribution to (\ref{CORR}) stems from the exchange of a closed string 
whose worldsheet spans a funnel between the two dipoles as first suggested by Veneziano in the context 
of the topological expansion~\cite{Veneziano:1976wm}.  In~\cite{Basar:2012jb}, (\ref{CORR}) was estimated using
a (quantum) closed string exchange 

\be
{\bf WW}=g_s^2\int_0^\infty\frac {dT}{2T}\,{\bf K}(T)\,,
\label{SCHWINGER}
\ee
with ${\bf K}(T)$ a string partition function with cylinder topology of modulus $T$ with twisted boundary conditions.
The string coupling $g_s\sim 1/N_c$  reflects on the subleading (handle contribution) character of the exchange
in large $N_c$.

The  dominant contribution to (\ref{SCHWINGER}) was found to follow from the
poles of the longitudinal stringy mode contributions  in~\cite{Basar:2012jb} ,

\be
{\bf WW}_{poles}&&\approx \frac {g_s^2}{4}
\sum_{k=1}^\infty\frac{(-1)^{k}}k \left(\frac{k\pi}{\chi}\right)^{D_\perp/2}\nonumber\\
&&\times\frac{aa'}{\alpha'} e^{-k{\bf b}_\perp^2/2\chi\alpha'+D_\perp\chi/12k}\, ,
\label{XPOLES}
\ee
with $\alpha'=1/(2\pi\sigma_T)$. The summation over $k$ is over the N-ality of the Wilson loop.
For  a loop in the fundamental representation $k=1$. This will  be understood below.
At large $\chi$ the string freezes in its ground state.
(\ref{XPOLES}) can be rewritten as
\be
{\bf WW}_{poles}&&\approx \frac {g_s^2}{4}\left(\frac{\pi}{\sigma_T}\right)^{{D_\perp}/2}\\
&&\times \sum_{k=1}^\infty\frac{(-1)^{k}}k\,\frac{aa'}{\alpha'}{\bf K}_k(\chi, {\bf b}_\perp)\nonumber \ .
\label{1XPOLES}
\ee
The emerging  propagator at the poles, 

\be
{\bf K}_k(\chi, {\bf b}_\perp) =\left(\frac{k}{2\pi\alpha'\,\chi}\right)^{D_\perp/2}\,e^{-k{\bf b}_\perp^2/2\chi\alpha'+D_\perp\chi/12k}\,,
\label{DIFF}
\ee
satisfies a diffusion equation in flat space 
\be
\left(\partial_\chi-\frac {D_\perp}{12k}\right)\,{\bf K}_k (\chi, {\bf b}_\perp) ={\bf D}_k\,\nabla^2_\perp\,{\bf K}_k(\chi, {\bf b}_\perp)\, ,
\label{DIFF1}
\ee
with a diffusion constant in rapidity space ${\bf D}_k=\alpha'/2k$.  For long strings with ${\bf b}_\perp$ large, the pertinent
AdS metric on the string is nearly flat. The effects of curvature will become apparent at intermediate ${\bf b}_\perp$ and will
be addressed below.

For long strings, the diffusion propagator (\ref{DIFF}) emerges as the natural version of the periodic string propagator in (\ref{SCHWINGER}) in the diffusive regime ${\bf b}_\perp\sim \sqrt{\chi\alpha'}$. 
We note that (\ref{DIFF1}) is just the proper time evolution of the tachyonic string mode 

\be
\left(\partial_{T_\perp}+(M_0^2-\nabla_\perp^2)\right)\,{\bf K}_k (T_\perp,M,{\bf b}_\perp)=0
\label{TACH2}
\ee
after the identification $T_\perp={\bf D}_k\chi$. The tachyonic mass follows from the harmonic string spectrum

\be
M_n^2=\frac{4}{\alpha'}\left(n-\frac{D_\perp}{24}\right)\rightarrow -\frac{D_\perp}{6\alpha'} \ .
\label{MASS2}
\ee

The occurence of (\ref{TACH2}) is naturally explained by noting that the dominant contribution to
the closed string propagator in (\ref{SCHWINGER})  stems from short proper times $T=2\pi k/\chi<1$. These short times
are equivalent to a large Unruh temperature $T_U\approx \chi/2\pi {\bf b}_\perp> 1$ on the string end-points~\cite{Basar:2012jb} . Because
of the open-closed string duality, this large Unruh temperature yields a dimensional reduction of the open string from
$D$ to $D_\perp$ dimensions. At large $\chi$ the string freezes to its lowest tachyonic mode and diffuses transversely.
The modular invariance at the origin of the open-closed string duality means that $T_\perp\sim 1/T\sim \chi$.
A large Unruh temperature means a large proper time evolution of the tachyonic string in proper time $T_\perp$.
This large proper time evolution is just the diffusion of the string in $D_\perp$ as first advocated by Gribov~\cite{GRIBOV}.

\section{Curved Tachyon Diffusion}

The preceding observations hold true even in curved space whereby the string in a curved background is also
subjected to a high Unruh temperature at the end points. Although the exact form of the string propagator in curved space
in (\ref{SCHWINGER}) is unknown, we expect the longitudinal pole structure leading to the reduction (\ref{XPOLES})
to remain unchanged since it follows from short proper times i.e.
$T\sim 1/\chi< 1$, which are insensitive to curvature. Since $\chi>1$ we still
expect a large Unruh temperature on the dual open string exchange, and therefore freezing and reduction of the 
string tachyon to transverse space which is curved on the diffusion time scale. Tachyon diffusion in curved space
follows through

\be
\left(\partial_{T_\perp}+(M_0^2-\frac 1{\sqrt{g_\perp}}\partial_\mu\,g_\perp^{\mu\nu}\sqrt{g_\perp}\,\partial_\nu)\right)\,\Delta_\perp (x_\perp, x'_\perp)=0 \ , \nonumber\\
\label{TACH3}
\ee
where we have suppressed ${\bf b}_\perp ,M$ to alleviate the notation. The metric  $g_\perp$ in (\ref{TACH3}) is that of the transverse
space with positive signature. (\ref{TACH3}) is the curved space generalization of (\ref{TACH2}).  Below, we show how
the curved diffusion propagator $\Delta_\perp$ can be substituted to ${\bf K}_k$  to generalize (\ref{XPOLES}-\ref{1XPOLES})
 to curved AdS.

In general $x_\perp$ is an arbitrary point in $D_\perp$.
In hyperbolic AdS type spaces it is useful to separate $x_\perp=({\bf x},z)$ with $z$ along the holographic direction
and ${\bf x}$ in the 2-dimensional physical space for $D_\perp=3$ for instance with diffusion in AdS$_3$. 	
The formal solution to (\ref{TACH3}) reads
\be
\Delta_\perp (x_\perp, x'_\perp)=<x_\perp|\,e^{-T_\perp(M_0^2-\nabla^2_C)}|x'_\perp> \ , 
\label{TACH4}
\ee
with $\nabla^2_C$ the curved Laplacian in (\ref{TACH3}) 
and $T_\perp={\bf D}\chi\gg 1$ for $k=1$. The transverse evolution 
propagator $\Delta_\perp$ in (\ref{TACH4})  ties to the tachyon propagator ${\bf G}(j)$

\be
<x_\perp| {\bf G}(j)\equiv \left({j+(M_0^2-\nabla_C^2)}\right)^{-1}|{x'}_\perp >
\label{GJ}
\ee
through an inverse  Mellin transform
\be
\Delta_\perp =\int_{\cal C}\frac{dj}{2i\pi}\, \,
{e^{jT_\perp}}\,{\bf G}(j) \ ,
\label{TACH5}
\ee
with ${\cal C}$ a pertinent contour in the complex $j$-plane at the rightmost of all singularities.  The 
tachyon propagator in (\ref{GJ})  obeys the curved equation
\be
\left(j+(M_0^2-\nabla_C^2)\right)\,{\bf  G}(j,x_\perp, x_\perp')=\frac{1}{\sqrt{g}}\,\delta_{D_\perp}(x_\perp-x_\perp') \ . \nonumber \\
\label{PRO1}
\ee
A similar propagator was noted in \cite{Brower:2006ea}  starting from the graviton 
using the critical closed string scattering amplitude in 10 dimensions.

\subsection{Conformal}

In transverse hyperbolic space AdS$_{D_\perp}$ with metric
\be
ds_\perp^2=\frac{1}{z^2}(d{\bf b}_\perp^2+dz^2) \ ,
\ee
all length scales are measured in units of the AdS radius which is set to 1, and reinstated
at the end by inspection.
The propagator for a scalar field is given by \cite{Berenstein:1998ij, Danielsson:1998wt}

\be
{\bf  G}_{D_\perp odd}(j,\xi)= \frac 1{4\pi} \left(\frac{-1}{2\pi \sinh(\xi) \frac{d}{d\xi}}\right)^{m-1} \frac{e^{-\nu \xi}}{\sinh(\xi)}  \label{PRO3}
\ee
for $D_{\perp}=2m+1$ and
\be
{\bf  G}_{D_\perp even}(j,\xi)= \frac 1{2\pi} \left(\frac{-1}{2\pi \sinh(\xi) \frac{d}{d\xi}}\right)^{m} \mathcal{Q}_{\nu-1/2}(\cosh(\xi)) \nonumber \\ \ 
\ee 
for $D_{\perp}=2m$ with 

\be
\nu^2=&&j-j_0\nonumber\\
j_0=&&-M_0^2-(D_\perp-1)^2/4 \ .
\ee
$\mathcal{Q}$ is a Legendre function of the second kind. The chordal distance $\xi$ is defined through
\be
{\rm cosh}\,\xi=1+d=1+\frac{{\bf b}_\perp^2+(z-z')^2}{2zz'} \ ,
\label{PRO4}
\ee
which gives for $\frac{{\bf b}_\perp^2}{2zz'}\gg 1$
\be
\xi \sim {\rm ln}\left(\frac{{\bf b}_\perp^2}{zz'}\right) \  , \ \ \ \ \sinh(\xi) \sim \frac{{\bf b}_\perp^2}{2zz'} \ . \label{PRO6}
\ee
For $D_\perp=3$, inserting the conformal propagator (\ref{PRO3}) in (\ref{TACH5}) yields 
the conformal evolution kernel
\be
\Delta_\perp(\chi,\xi)=\frac{e^{j_0 {\bf D} \chi}}{(4 \pi {\bf D} \chi)^{3/2}} \frac{\xi e^{-\frac{\xi^2}{4{\bf D} \chi}}}{\sinh(\xi)} \ ,
\label{PRO5}
\ee
with the diffusion constant ${\bf D}=\alpha'/2=1/(2 \sqrt \lambda)$ for conformal AdS$_3$,
after restricting the N-ality to $k=1$. This heat kernel was obtained in \cite{David:2009xg} using a group theoretical approach. (\ref{PRO5})
shows that in conformal AdS$_3$, the tachyon mode of the bosonic string diffuses in hyperbolic space
along the geodesic distance as measured by $\xi$ which is about twice the chordal distance for small
displacement, i.e. $\xi^2\approx 2d\ll 1$. Again the rapidity $\chi$ plays the role of time. \\

\subsection{Confining}

Confinement in AdS is captured in a simple way by the hard-wall model, whereby only a slice of the AdS space is considered 
with $0\leq z\leq z_0$ and $z_0\approx 1/\Lambda$  setting up the confinement scale~\cite{Polchinski:2002jw}. In this case,
all scales are set by $z_0$ implicitly in the intermediate expressions and explicitly in the final ones. We note that in the hard
wall model we still use the identification $\alpha'\equiv l_s^2/z_0^2\equiv\sqrt{\lambda}$.

To simplify the  analysis for the curved diffusion, we define the total {\it wee-dipole} density ${\bf N}=\Delta/(zz')^{D_\perp-2}$. Since the scattering amplitude is symmetric under the interchange of the two dipoles and we are going to identify $z, z'$ with the effective size of the dipoles, the correct rescaling of ${\bf N}$ is by powers of $z z'$. 
Using the conformal variable $u=-{\rm ln}(z/z_0)$, the diffusion equation for the dipole density reads
\be
\left(\partial_{T_\perp}+(M_0^2+D_\perp-2)-\partial_u^2-e^{2u}\nabla^2_{{\bf b}_\perp}\right)\,{\bf N}=0 \ ,
\label{HW1}
\ee
The proper time evolution of ${\bf N}$ in AdS amounts to a transport or diffusion equation,
with the initial condition
\be
{\bf N}(T_\perp=0,u,u',{\bf b}_\perp) =\delta(u-u')\delta({{\bf b}}_\perp)  
\label{HW2}
\ee
as one-dipole per unit area in the transverse ${\bf b}_\perp$.

The boundary condition for solving (\ref{HW1}) follows from the conservation of the diffusion charge
in the slab $0\leq z\leq z_0$ or $0\leq u\leq \infty$,

\be
&&\frac d{dT_\perp}\,\int\,du\,d{\bf b}_\perp \,e^{T_\perp(M_0^2+D_\perp-2)}\,{\bf N} \nonumber \\
&& \ \  \ = \int d{\bf b}_\perp e^{T_\perp(M_0^2+D_\perp-2)}\,\partial_{u=0}\,{\bf N} \ ,
\label{HW3}
\ee
assuming that the diffusion current vanishes at $|{\bf b}_\perp|=\infty$  and at $u=\infty$ or the 
UV boundary condition as no holographic source is subsumed. Thus, the Neumann boundary
condition 

\be
\partial_{u=0}{\bf N}=0 
\label{NEU}
\ee
enforces that the (singlet) {\it wee-dipole}   current does not leak in the infrared at $z=z_0$.
As a result 

\be
\int\,du\,d{\bf b}_\perp \,e^{T_\perp(M_0^2+D_\perp-2)}\,{\bf N} \equiv 1
\label{NORM}
\ee
is fixed both in the conformal and confining case.
Other boundary conditions on the wall, e.g. absorptive or mixed, will resut in {\it wee-dipole}  
current loss in the infrared or confining region, with  (\ref{NORM}) less than 1.
While intricate physically, these boundary conditions will be pursued  elswhere.

The solution to (\ref{HW1}) subject to (\ref{HW2}-\ref{NEU}) is readily obtained by the image 
method for the current conserving Neumann boundary condition
\be
{\bf N}(T_\perp,u,u',{\bf b}_\perp)&=&\frac{1}{z_0^2}e^{u'+u}\,\Delta(\chi,\xi)+\frac{1}{z_0^2}e^{u'-u}\,\Delta(\chi,\xi_*)  \nonumber \\ 
&=& \frac{1}{zz'}\,\Delta(\chi,\xi)+\frac{z}{z'z_0^2}\,\Delta(\chi,\xi_*) \label{HW5} \ ,
\ee
with the conformal solution (\ref{PRO5}) for $\Delta (\chi,\xi)$. The invariance of the interchange of the two dipoles in the conformal case gets affected in the confining contribution (second part in (\ref{HW5})). 
The chordal distances follow from (\ref{PRO4}) as
\be
{\rm cosh}\xi&=&{\rm cosh}(u'-u)+\frac 12 {\bf b}_\perp^2\,e^{u'+u}\, \\
{\rm cosh}\xi_*&=&{\rm cosh}(u'+u)+\frac 12 {\bf b}_\perp^2 e^{u'-u}\,\nonumber \ ,
\label{HW6}
\ee
with $-u$ the image of $u$ with respect to the holographic wall at $u=0$.

\subsection{Wee-Dipole Density}

${\bf N}$ obeys a "Markovian" type chain rule
\be
\int\,du" d{\bf b}_\perp"
&&{\bf N}(T_\perp-T_\perp", u,u",{\bf b}_\perp-{\bf b}_\perp")\,\nonumber\\
\times&&{\bf N}(T_\perp"-T_\perp', u",u',{\bf b}_\perp"-{\bf b}_\perp')\,\nonumber\\
=&&{\bf N}(T_\perp -T_\perp', u,u',{\bf b}_\perp-{\bf b}'_\perp) \ ,
\label{HW10}
\ee
which follows readily from the diffusion evolution kernel as a propagator in rapidity space.
(\ref{HW10}) suggests a "Weizsaecker-Williams" analogy for the virtual dipole field surrounding each
of the initial projectile and target dipole.
Thus, the total number of {\it wee-dipoles} either in the target or the projectile follows from the normalization
\be
N_{\it wee}=\int\,du\,d{\bf b}_\perp \,{\bf N} =e^{-T(M_0^2+D_\perp -2)} \equiv (s/s_0)^{\alpha_{\bf P}-1} \, 
\label{HW4}
\ee
with the $1/\sqrt{\lambda}$ corrected intercept

\be
\alpha_{\bf P}=1+\frac{D_\perp}{12} - \frac{(D_\perp-1)^2}{8 \sqrt{\lambda}} \ .
\ee
${\bf N}$ is interpreted as the density of {\it wee-dipoles} of scale $u$ at a transverse
distance ${\bf b}_\perp$ sourced by a dipole of scale $u'$ located at ${\bf b}'_\perp={\bf 0}$. Their
total number or multiplicity  is given by (\ref{HW4}) and grows exponentially with the rapidity 
$\chi=T/{\bf D}\equiv{\rm ln}(s/s_0)$. This growth is at the origin of the violation of unitarity in
the scattering amplitude. Here it is tamed by the eikonalized amplitude whereby a class of
$1/N_c$ corrections are resummed.

\begin{figure}[t]
  \begin{center}
  \includegraphics[width=8cm]{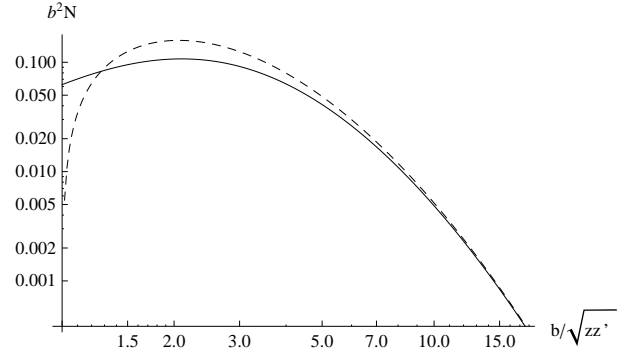}
  \caption{Holographic {\it wee-dipole} spatial distributions for $z=z'=1.8\,GeV^{-1}$ and $\chi=10$.  
  Confining density in (\ref{HW5}): solid curve; asymptotic density in (\ref{WEEX}): dashed curve. See text.}
    \label{densitycomp}
  \end{center}
\end{figure}

\begin{figure}[t]
  \begin{center}
  \includegraphics[width=8cm]{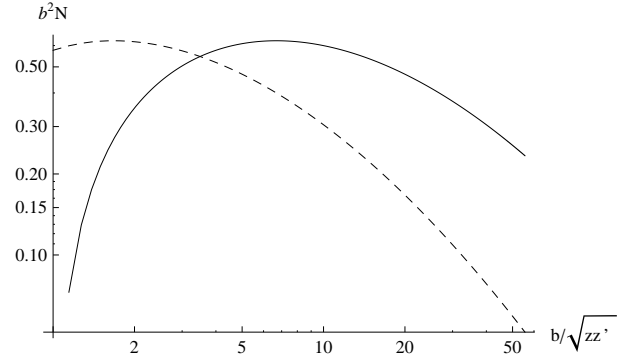}
  \caption{BFKL {\it wee-dipole} spatial distributions for $z=z'=1.8\,GeV^{-1}$ and $\chi=10$.  
  BFKL density (\ref{DBFKL}): solid curve; improved BFKL density: dashed curve. See text.}
    \label{densitycompbfkl}
  \end{center}
\end{figure}

Using the chain rule (\ref{HW10}) and the dipole density (\ref{HW5}), we obtain the asymptotic dipole density

\be
{\bf N}(\chi, z,z', {\bf b}_\perp) \approx  2 \frac{e^{(\alpha_{\bf P}-1) \chi}}{\left(4\pi {\bf D} \chi\right)^{3/2}} \frac{z }{z'{\bf b}_\perp^2} {\rm ln}\left(\frac{{\bf b}_\perp^2}{z z'}\right) e^{-{\rm ln}^2\left(\frac{{\bf b}_\perp^2}{z z'}\right)/(4{\bf D} \chi)}  \nonumber \\
\label{WEEX}
\ee
in the conformal case and in the limit  $\frac{{\bf b}_\perp^2}{2zz'}\gg 1$. 
The analogue of (\ref{WEEX}) in the
context of onium-onium scattering was discussed in~\cite{Mueller:1993rr,Mueller:1994jq,Mueller:1994gb}. In particular, in the
BFKL 1-pomeron approximation it is  given by~\cite{Mueller:1994gb} 

\be
&&{\bf N}^{\bf BFKL}(\chi, z,z', {\bf b}_\perp) \approx 2 \frac{e^{(\alpha^{\bf BFKL}-1)\chi}}{\left(4\pi {\bf D}^{\scriptscriptstyle \bf BFKL} \chi \right)^{3/2}} \label{DBFKL} \\
&& \ \ \ \ \ \ \ \ \ \ \ \ \times\frac{z}{z'{\bf b}_\perp^2}{\rm ln}\left(\frac{{\bf b}_\perp^2}{zz'} \right) e^{-{\rm ln}^2\left(\frac{{\bf b}_\perp^2}{zz'}\right)/(4 {\bf D}^{ \bf BFKL} \ \chi)} \ ,
\nonumber
\ee 
with the BFKL intercept $\alpha^{\bf BFKL}$ and diffusion constant ${\bf D}^{\scriptscriptstyle \bf BFKL}$, see (\ref{onium}).  Modulo 
the pomeron intercept and the diffusion constant, which are different (weak coupling or BFKL versus strong coupling or  holography), 
the holographic result in the conformal limit is identical to the BFKL 1-pomeron approximation.
Again, the occurence of the $3/2$ exponent reflects on diffusion in $D_\perp=3$ as noted earlier.
It is remarkable that the BFKL resummation of perturbative QCD diagrams is encoded in the stringy Schwinger
mechanism discussed in~\cite{Basar:2012jb}, albeit in hyperbolic  space.

Figure~\ref{densitycomp} shows the distribution of the holographic  {\it wee-dipole} density (\ref{HW5}) in solid line 
versus ${\bf b}_\perp$ for $z=z'=1.8\,GeV^{-1}$ and $\chi=10$. The dashed curve is the asymptotic distribution (\ref{WEEX}).
Figure~\ref{densitycompbfkl} shows the distribution of the BFKL {\it wee-dipole}  density (\ref{DBFKL}) in solid line
versus ${\bf b}_\perp/\sqrt{zz'}$ for also $z=z'=1.8\,GeV^{-1}$ and $\chi=10$. The dashed curve is the ÒimprovedÓ BFKL result in
\cite{Salam:1995uy}. The latter follows from (\ref{DBFKL}) by inserting a factor of 16 in the argument of the logarithm
which corresponds to rescaling down the BFKL distribution by a factor of 4 along the ${\bf b}_\perp/\sqrt{zz'}$ axis.
The holographic results use: ${\bf D}=0.10$ and ${\alpha}_{\bf P}-1=0.146$, while the BFKL results use:
${\bf D}^{\bf BFKL}=0.72$ and $\alpha_{\bf P}-1=0.477$ with $\lambda=23$. Both the holographic and the improved distributions
are less skewed and more centered. The holographic distribution is less spread than the improved BFKL distribution
in ${\bf b}_\perp$, therefore less infrared sensitive.

\section{Saturation in curved space}

In curved space the holographic picture suggests the identification of the holographic direction with the effective size of the scatterer, \cite{Polchinski:2001tt, Polchinski:2002jw,Brower:2010wf, Brower:2011dx}. We now suggest that dipole-dipole scattering in holography can be thought of as scattering a {\it wee-dipole} of virtuality  $1/z$ onto a
{\it wee-dipole} of virtuality $1/z'$. This leads to the concept of dipole saturation as first discussed 
in~\cite{Gribov:1984tu}.

For $D_\perp=3$, we identify
\be
\frac{aa'}{\alpha'}\,{\bf K}_k(\chi, {\bf b}_\perp)\rightarrow z z' \ {\bf N}(\chi,z,z',{\bf b}_\perp) \ .
\label{EQ1}
\ee
In terms of (\ref{EQ1}), the leading ($k=1$) contribution to (\ref{XPOLES}) in a curved AdS background reads
\be
{\bf WW}_{poles} \approx - \frac{g_s^2}{4} \left(2\pi \alpha' \right)^{3/2} z z' {\bf N}(\chi, z,z', {\bf b}_\perp) \ . 
\label{WWXX}
\ee
The differential dipole-dipole cross section at finite impact parameter is then~\cite{Basar:2012jb}

\be
\frac{d^4 \sigma_{tot}}{du du' d{\bf b}_\perp} = 2 (1-e^{{\bf WW}_{poles}}) \label{crosssection1}  \ .
\ee
Assuming that the target is a proton with a dipole wave function peaked at some virtuality corresponding to $u_{\bf T}$, i.e. ${\varphi}_{\bf T}(u')=\delta(u'-u_{\bf T})$, (\ref{crosssection1}) averaged over a target wave function reads 
\be
\frac{d^3\sigma_{\rm tot}}{du \, d{\bf b}_\perp}\approx 2\left(1-e^{\left<{\bf WW}_{\rm poles}\right>}\right) \ ,
\label{SAT3}
\ee
with only the first cumulant retained and
\be
\left<{\bf WW}_{\rm poles}\right>=\int \,dz\,\varphi_{\bf T}(z)\,{\bf WW}_{\rm poles} \ .
\label{SAT4}
\ee
Higher cumulants are suppressed by higher powers of $g_s^2\approx 1/N_c^2$. 
This amounts to $z'\rightarrow z_{\bf T}$ in (\ref{WWXX}).
\begin{figure}[t]
  \begin{center}
  \includegraphics[width=8cm]{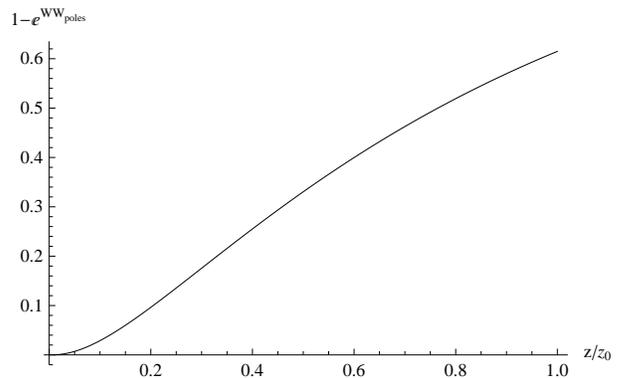}
  \caption{Saturating behavior of the dipole-dipole cross section, (\ref{SAT3}), at $\chi=15$ for fixed
 $ {\bf b}_\perp=2 GeV^{-1}$. See text.}
  \label{satline}
  \end{center}
\end{figure}

(\ref{SAT3}) suggests the definition of the saturation momentum ${\bf Q}_s$ from
\be
\frac{d^3\sigma_{\rm tot}}{du_s \, d{\bf b}_\perp}\equiv 2\left(1-e^{-z_s {\bf Q}_s/(2\sqrt{2})}\right) \ ,
\label{SAT5}
\ee
with ${\bf Q}_s\equiv - 2 \sqrt{2}\left<{\bf WW}_{\rm poles}\right>/z_s$ fixed by the saturating dipole size
$z_s=\sqrt{2}/{\bf Q}_s$. This saturating behavior is illustrated in Figure \ref{satline}. This is the canonical choice for which

\be
\left(1-e^{-z_s{\bf Q}_s/(2\sqrt{2})}\right)\rightarrow \left(1-e^{-1/2}\right)\approx 0.4 \ ,
\ee
leading to a scattering amplitude of order 1.  The saturation momentum follows from the transcendental equation

\be
\frac{z_s}{\sqrt{2}} {\bf Q}_s (\chi , {\bf b}_\perp) 
= \frac{g_s^2}{2} \left(2\pi \alpha' \right)^{3/2} \, z_sz_{\bf T}\,{\bf N}(\chi, z_s,z_{\bf T}, {\bf b}_\perp) =1  .  \nonumber \\
\label{transzendental} 
\ee

\begin{figure}[t]
  \begin{center}
  \includegraphics[width=8cm]{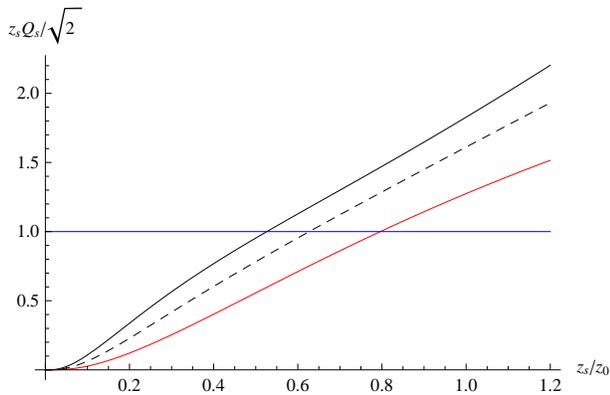}
  \caption{Illustration of the solutions to (\ref{transzendental}) with ${\bf b}_\perp^2=0 \ GeV^{-2}$ (black, solid), ${\bf b}_\perp^2=1 \ GeV^{-2}$ (black, dashed) and ${\bf b}_\perp^2=3 \ GeV^{-2}$ (red) at $\chi=8$. See text.}
  \label{Qsatzs}
  \end{center}
\end{figure}
Saturation takes place whenever the dipole density ${\bf N}\sim N_c^2/\lambda^{\frac 54}> 1$ in (hard wall) holography. 
This is comparable to perturbative QCD with ${\bf N}\sim N_c^2/\lambda> 1$. As the dipole density ${\bf N}(\chi, z_s,z_{\bf T}, {\bf b}_\perp)$ is peaked around some finite $z_s$ for fixed $\chi,z_{\bf T}, {\bf b}_\perp$, the solution to (\ref{transzendental}) has in general two solutions. To explicit them, we now need to detail the holographic parameters.

Our set of dimensionless holographic parameters consists of:
$D_\perp=3$, $N_c=3$, $\lambda=23$ and $\kappa=2.5$. 
The choice of $\lambda$ is fixed by the $F_2$ slope in
comparison to the DIS data (see below). The value of $\kappa$ 
is fixed by the saturation scale (see below).
Since $\lambda=g^2N_c$, the Yang-Mills coupling is
$g^2/4\pi=0.6$, which is on the strong coupling side.

We note that 
although the {\it original} string coupling is small, i.e. ${\lambda}/4\pi N_c=0.6<1$ as required by holography, 
the physical value of $N_c<\lambda$ is at odd with the holographic and strong coupling  limit. This notwithstanding,
our corrected soft pomeron intercept is

\be
\alpha_{\bf P}-1=\frac 14 -\frac{1}{2\sqrt{\lambda}}=0.146 \ .
\ee
Although this numerical value is on the higher side of the $pp$ scattering data of 0.08, it only refers to the 
{\it bare} soft pomeron intercept which is likely to decrease through multipomeron resummation and
shadowing.

Our set of dimensionfull holographic parameters consists of:
$z_0=2 \ GeV^{-1}$, $z_{\bf T}=1.8 \ GeV^{-1}$, $s_0=10^{-1} \ GeV^{2}$,
which are set close to the confining scale in QCD. We kinematically
translate the rapidity through

\be
\chi={\rm ln} \left(\frac{s}{s_0}\right)\equiv{\rm ln} \left(\frac{Q^2}{s_0}\left(\frac 1x -1\right)\right)\nonumber
\ee
using the DIS kinematics (see below).
For fixed $\chi=5$ and $z_{\bf T}=2$ and varying ${\bf b}_\perp$, 
an illustration of (\ref{transzendental}) is shown in Figure \ref{Qsatzs}, using these parameters. The numerical
dependence of the slope in Figure \ref{Qsatzs} near the origin (small $z_s/z_0$) is linear.

\subsection{Conformal}

In the conformal limit, the dipole density is explicit,  giving the implicit saturation density

\be
{\bf Q}_s (\chi ,{\bf b}_\perp)=  \frac{g_s^2}{\sqrt{2}} \left(2\pi \alpha' \right)^{3/2} \frac{1}{z_s} \frac{e^{(\alpha_{\bf P}-1) \chi}}{\left(4 \pi {\bf D} \chi\right)^{3/2}}  \frac{\xi e^{-\frac{\xi^2}{4{\bf D} \chi}}}{\sinh(\xi)} \ . \label{SAT6} 
\ee
For large transverse separation $\frac{{\bf b}_\perp^2}{2z_sz_{\bf T}}\gg 1$,  (\ref{SAT6})
defines a dipole density in the transverse coordinate

\begin{figure}[!tbp]
  \begin{center}
  \includegraphics[width=8cm]{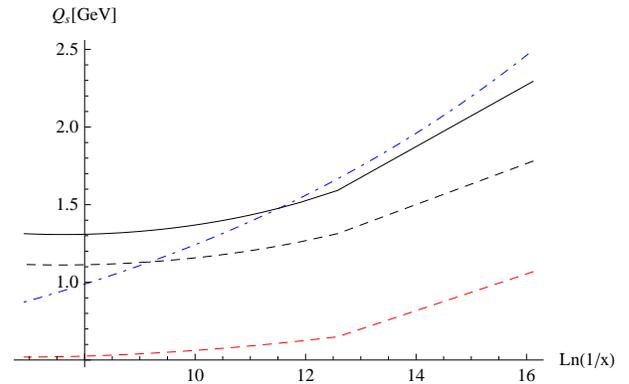}
  \caption{$x$-dependence of the saturation momentum (\ref{SAT8}). Black, solid: ${\bf b}_\perp^2=0 \ GeV^{-2}$. Black, dashed: ${\bf b}_\perp^2=1 \ GeV^{-2}$. Dashed red curve: saturation momentum in the conformal limit, (\ref{SAT6}), with ${\bf b}_\perp^2=1 \ GeV^{-2}$. The dashed dotted blue curve is the GBW saturation momentum from (\ref{GBW}). See text.} 
  \label{Qsatx}
  \end{center}
\end{figure} 

\begin{figure}[!tbp]
  \begin{center}
  \includegraphics[width=8cm]{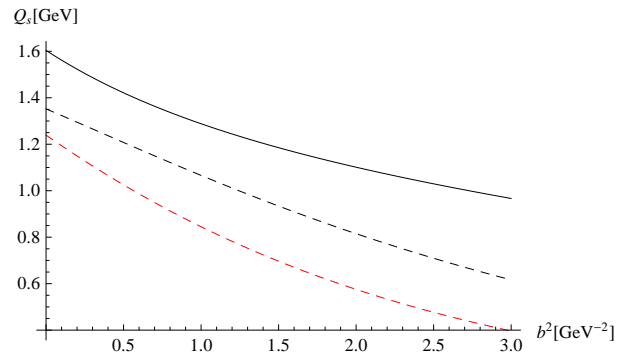}
  \caption{Impact parameter dependence of the saturation momentum (\ref{SAT8}). Upper curve (black, solid): $\chi=5$, lower curve (dashed, solid): $\chi=8$. Lowest curve (red, dashed): Saturation momentum in the conformal limit, (\ref{SAT6}), $\chi=8$. See text.}
  \label{Qsatb}
  \end{center}
\end{figure}

\be
{\bf Q}_s (\chi ,{\bf b}_\perp)
&\approx&   \sqrt{2}g_s^2 \left(2\pi \alpha' \right)^{3/2} \frac{e^{(\alpha_{\bf P}-1)\chi}}{\left(4 \pi {\bf D} \chi\right)^{3/2}} \nonumber \\
& \times& \frac{z_{\bf T}}{{\bf b}^2_\perp}\,{\rm ln}\left(\frac{{\bf b}_\perp^2}{z_sz_{\bf T}}\right) e^{-{\rm ln}^2\left(\frac{{\bf b}_\perp^2}{z_sz_{\bf T}}\right)/(4{\bf D} \chi)} \ .
\label{SAT66}
\ee
The large $\chi={\rm ln} \left({s}/{s_0}\right)> 1$ exponential asymptotics of (\ref{SAT6}-\ref{SAT66}) have two solutions,
say $z_{s1}<z_{s2}$. Only the small dipole solution $z_{s1}$ is retained in the conformal case, as the large dipole solution
$z_{s2}$ is deep in the infrared and unphysical. In the confined case, it is naturally cutoff by the wall (see below). With this
in mind and to leading exponential accuracy

\be
{\bf Q}_s(\chi , {\bf b}_\perp)\approx
\frac{z_{\bf T}}{{\bf b}^2_\perp}\,e^{2{\bf D}\chi\left(\sqrt{1+(\alpha_{\bf P}-1)/{\bf D}}-1\right)} \ .
\label{QASYMP}
\ee
At large $\sqrt{\lambda}$

\be
{\bf Q}_s(\chi , {\bf b}_\perp)\approx
\frac{z_{\bf T}}{{\bf b}^2_\perp}\,\left(\frac 1x\right)^{\sqrt{D_\perp/6\sqrt{\lambda}}} \ ,
\ee
illustrating the smallness of the exponent. For the parameters used above, (\ref{QASYMP}) reads

\be
{\bf Q}_s(\chi , {\bf b}_\perp)\approx
\frac{z_{\bf T}}{{\bf b}^2_\perp} \left(\frac{1}{x}\right)^{0.228/2} \ .
\label{ASYMP}
\ee
An early phenomenological approach to describe DIS data at HERA by Golec-Biernat and Wuesthoff
(GBW) in~\cite{GolecBiernat:1998js}, defines the saturation momentum as 
\be
Q_s^{\scriptscriptstyle \bf GBW}(x) = \left( \frac{x_0}{x} \right)^{\Lambda/2} \ GeV \ . \label{GBW}
\ee
HERA data are fitted with $x_0=3.04 \ 10^{-4}$ and $\Lambda=0.288$. Note that the GBW saturation momentum 
corresponds to the substitution $z_s{\bf Q}_s/(2\sqrt{2})\rightarrow (z_sQ_s^{\bf GBW}/2)^2$ in (\ref{SAT5}).
While the magnitude of the saturation momentum in our holographic approach can be adjusted by tuning $\kappa$, we find that the $x-$dependence of the saturation momentum, (\ref{ASYMP}), agrees well with the phenomenological fit in (\ref{GBW}) as shown in Figure \ref{Qsatx}.

\subsection{Confining} 

The identification (\ref{EQ1}) carries over to the confining case. The saturation momentum follows from the transcendental equation
\be
&&{\bf Q}_s (\chi ,{\bf b}_\perp) =  \frac{g_s^2}{2} \left(2\pi \alpha' \right)^{3/2}\nonumber\\
&&\times  \frac{e^{(\alpha_{\bf P}-1) \chi}}{\left(4 \pi {\bf D} \chi\right)^{3/2}}  
\Big( \frac{1}{z_s}  \frac{\xi e^{-\frac{\xi^2}{4{\bf D} \chi}}}{\sinh(\xi)} 
+\frac{z_s}{z_0^2}  \frac{\xi_* e^{-\frac{\xi_*^2}{4{\bf D} \chi}}}{\sinh(\xi_*)}  \Big) \ .
\label{SAT8}
\ee
The $x$-dependence of the saturation momentum (\ref{SAT6}),(\ref{SAT8}) is shown in Figure \ref{Qsatx}. Figure \ref{Qsatb} shows the relevant solution ($z_s \le z_0$) for the saturation momentum. Note the slow dependence of the holographic saturation momentum
on the longitudinal energy in the range ${\rm ln}(1/x)\leq 12$. Also 
note the non-trivial dependence on the impact parameter in the scattering amplitude or $Q_s$ as opposed to a factorization approach done in most saturation and color glass-condensate models, compare \cite{Tribedy:2010ab} and references within.

\section{DIS in Curved Space}

DIS of a lepton on a proton target can be viewed as a small size dipole scattering through a proton
\cite{Mueller:1989st, Mueller:1994gb,Mueller:1993rr,Mueller:1994jq,Iancu:2003uh, Nikolaev:1990ja,Nikolaev:1991et}.
Dipole-dipole scattering using Wegner-Wilson loops to discribe high-energy reactions of hadrons and photons was discussed
in~\cite{Nachtmann:1996kt,Shoshi:2002in}. A holographic approach to DIS starting from the graviton limit and based on the critical string amplitude was elaborated in~\cite{Brower:2010wf, Brower:2011dx}, see also \cite{Kovchegov:2009yj}.  Our approach is non-critical and rooted in the dipole-dipole formulation as detailed in~\cite{Basar:2012jb}  and briefly reviewed above. 

The dipole-dipole cross section is useful for the determination of the inclusive proton structure function $F_2(x,Q^2)$ for
small Bjorken $x$ and large $Q^2$. Specifically~\cite{Nikolaev:1990ja,Nikolaev:1991et},
\be
F_2(x,Q^2)=\frac{Q^2}{4\pi^2\alpha_{EM}}\,\left(\sigma_T+\sigma_L\right) \ ,
\label{F2}
\ee
and $\sigma_T+\sigma_L=\sigma_{\rm tot}$ can be regarded as the total (virtual) photon-to-proton 
or dipole-to-dipole cross section. By the optical
theorem 
\be
\sigma_{\rm tot}(s)=-\frac 1s\,{\rm Im}\, {\cal T}(s,t=0) \ ,
\label{OPTICAL}
\ee
whereby
\be
{\cal T}(s,0)=-2is\int d{\bf b}_\perp du du'\,{\varphi}_{\bf P}(u){\varphi}_{\bf T}(u') \,
\left(1-e^{\bf WW_{\rm poles}}\right), \nonumber\\ 
\label{DD}
\ee
which is an averaging of the $zz'$-dipole-dipole cross-section over the target $\varphi_{\bf T}(u')$ and
projectile ${\varphi }_{\bf P}(u)$ dipole wave functions respectively. Thus,
\be
&&F_2(x,Q^2)=\nonumber\\
&&\frac{Q^2}{2\pi^2\alpha_{EM}}\int\,d{\bf b}_\perp
\,du\,du'\,{\varphi}_{\bf P}(u){\varphi}_{\bf T}(u')\,
\left(1-e^{\bf WW_{\rm poles}}\right) \ . \nonumber \\
\label{DD1}
\ee
Typically, the (target) proton and (projectile) photon dipole distributions are peaked, say
\be
\varphi_{\bf P}(u)\equiv &&\left(\alpha_{EM} / \kappa^2\right) \delta(u-u_{\bf P})\nonumber\\
\varphi_{\bf T}(u')\equiv && \delta(u-u_{\bf T}) \ .
\label{DD2}
\ee
The normalization of the projectile (current) distribution in (\ref{DD2}) is fixed empirically by the magnitude of the
measured structure function $F_2$.

\begin{figure}[!htbp]
  \begin{center}
  $\begin{array}{ll}
  \includegraphics[width=4.2cm]{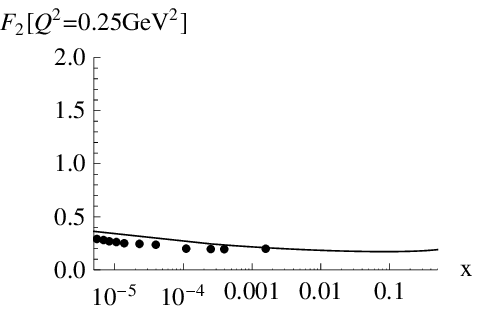} & \includegraphics[width=4.2cm]{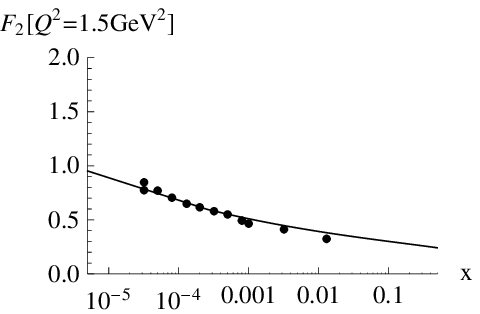} \nonumber \\
  \includegraphics[width=4.2cm]{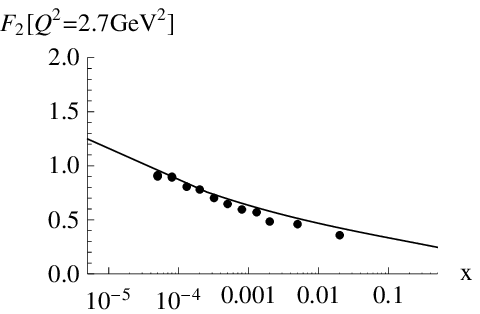} & \includegraphics[width=4.2cm]{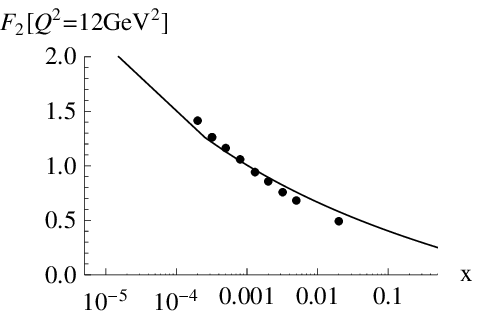} \nonumber \\
  \includegraphics[width=4.2cm]{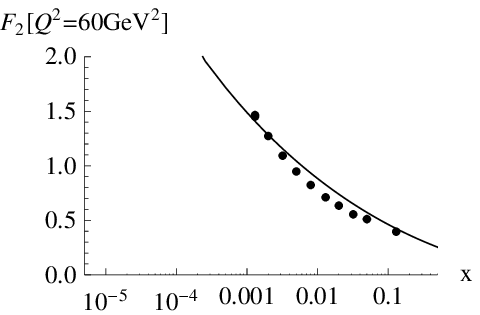} & \includegraphics[width=4.2cm]{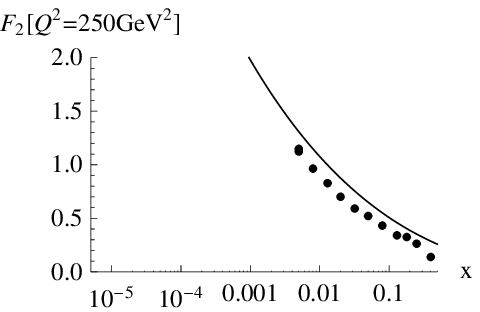} \nonumber \\
  \includegraphics[width=4.2cm]{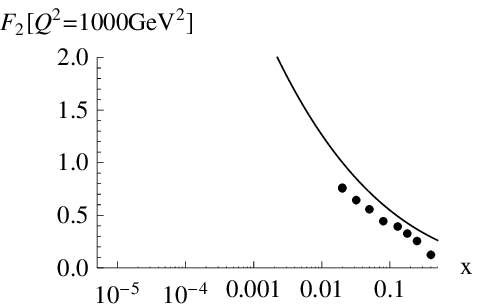} & \includegraphics[width=4.2cm]{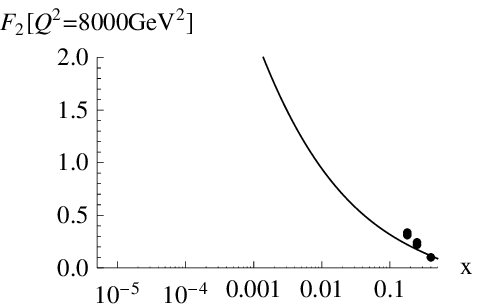} \nonumber \\
  \end{array}$
  \caption{Proton structure function $F_2$ in the confining background, (\ref{DD5}). See text.}
  \label{F2confining}
  \end{center}
\end{figure}
\begin{figure}[!htbp]
  \begin{center}
  $\begin{array}{ll}
  \includegraphics[width=4.2cm]{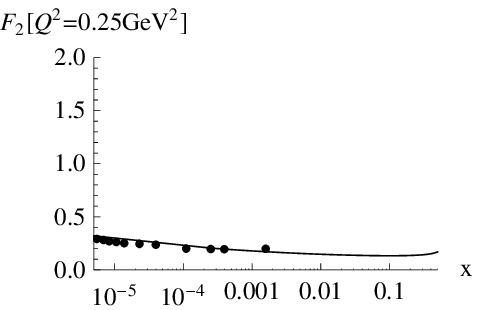} & \includegraphics[width=4.2cm]{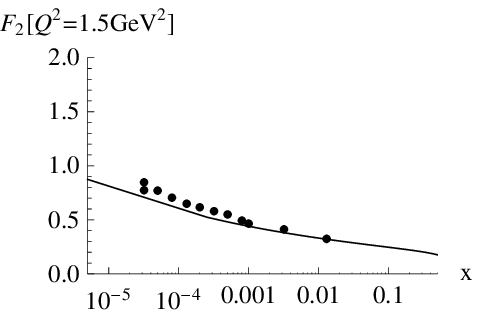} \nonumber \\
  \includegraphics[width=4.2cm]{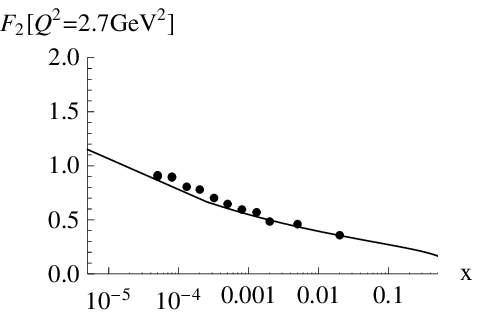} & \includegraphics[width=4.2cm]{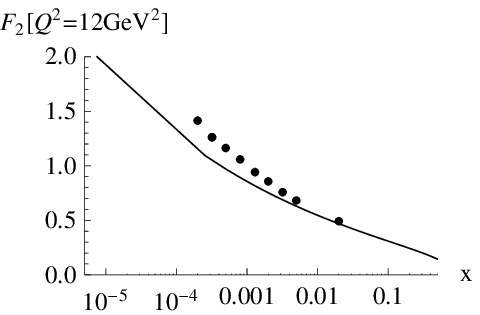} \nonumber \\
  \includegraphics[width=4.2cm]{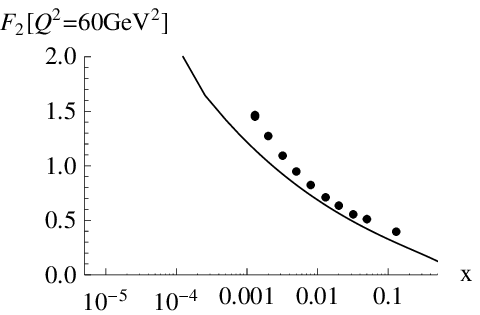} & \includegraphics[width=4.2cm]{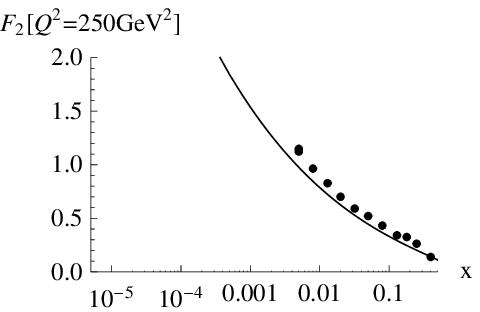} \nonumber \\
  \includegraphics[width=4.2cm]{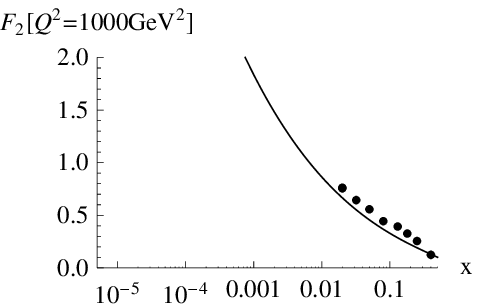} & \includegraphics[width=4.2cm]{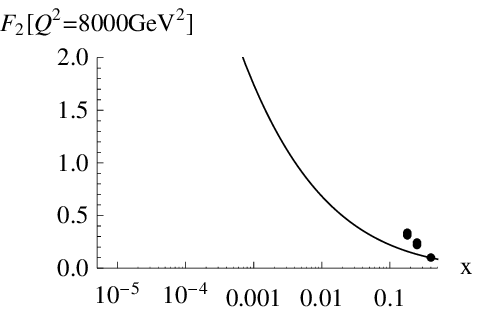} \nonumber \\
  \end{array}$
  \caption{Proton structure function $F_2$ in the conformal limit, (\ref{DD4}). See text.}
  \label{F2conformal}
  \end{center}
\end{figure}
Inserting (\ref{DD2}) into (\ref{DD1}) and using the dipole-dipole cross-section
for $D_\perp=3$ yields in the 1-pomeron exchange limit  and in the conformal case
\be
&&F_2(x,Q^2)\Big|_{\rm conformal}\approx  
\frac{g_s^2}{8 \pi^2 \kappa^2} \left(2\pi \alpha' \right)^{3/2}  \nonumber\\
&&\times  z_{\bf T} \, Q \frac{e^{(\alpha_{\bf P}-1)\chi}}{\sqrt{4\pi{\bf D}\chi}}
\,\left(e^{-\frac 1{4{\bf D}\chi}{\rm ln}^2(Qz_{\bf T})}\right)
\label{DD4}
\ee
and in the confining case
\be
&&F_2(x,Q^2)\Big|_{\rm confining}\approx  \frac{g_s^2}{8 \pi^2 \kappa^2} \left(2\pi \alpha' \right)^{3/2} \nonumber\\
&&\times z_{\bf T} \  Q \frac{e^{(\alpha_{\bf P}-1) \chi}}{\sqrt{4\pi{\bf D}\chi}}
\,\left(e^{-\frac 1{4{\bf D}\chi}{\rm ln}^2(Qz_{\bf T})}+e^{-\frac 1{4{\bf D}\chi}{\rm ln}^2(Qz_0^2/z_{\bf T})}\right)\nonumber \ , \\ 
\label{DD5}
\ee
with $u_{\bf P}={\rm ln}(z_0Q)$, $u_{\bf T}={\rm ln}(z_0/z_{\bf T})$. We have used the fact that
\be
&&{\bf N}(T_\perp,u,u',t=0)= \nonumber \\
&&\frac{e^{-T_\perp(M_0^2+1)}}{\sqrt{4\pi T_\perp}} 
\left(e^{-(u'-u)^2/4 T_\perp}+e^{-(u'+u)^2/4T_\perp}\right)
\label{HW8}
\ee
for $t=-{{\bf q}}^2_\perp=0$, after making use of the Fourier transform

\be
{\bf N}(T_\perp,u,u',{{\bf q}}^2_\perp)=\int d{\bf b}_\perp \,e^{i{{\bf q}}_\perp\cdot {\bf b}_\perp} \, {\bf N}(T_\perp,u,u',{\bf b}_\perp) \ .
\label{HW7}
\ee
Since the diffusion kernel in (\ref{HW8}) is generic, the $Q^2$ dependency of the structure factor is sensitive to the $\perp$-dimensions considered,  ${\bf N}=\Delta/(zz')^{D_\perp-2}$. 

The above approximation is justified when the photon momentum is sufficiently larger than the saturation scale, $Q \ge Q_s$, at all impact parameters ${\bf b}_\perp$. For the range of values for $Q^2$, $x$ considered to compare to the HERA data, the value for the proton structure function $F_2$ using the exponentiated 1-pomeron exchange, (\ref{DD1}), differs by less than $8\%$ compared to $F_2$ from (\ref{DD5}).

Figures \ref{F2confining}, \ref{F2conformal} compare our results (\ref{DD5}), (\ref{DD4}) 
to the HERA data~\cite{Aaron:2009aa}, using the holographic parameter set quoted above.
Figure \ref{F2conformal} shows that with our choice of parameters, our result for the 1-pomeron exchange amplitude in the 
conformal and confining metric appears to fit equally well the DIS data overall. A closer scrutiny of the fits show a better 
fit by the confining metric at low $Q^2$. Clearly our analysis is only qualitative, and a more thorough
study of the parameter dependences and the fitting accuracy are needed.

\begin{figure}[t]
  \begin{center}
  \includegraphics[width=8cm]{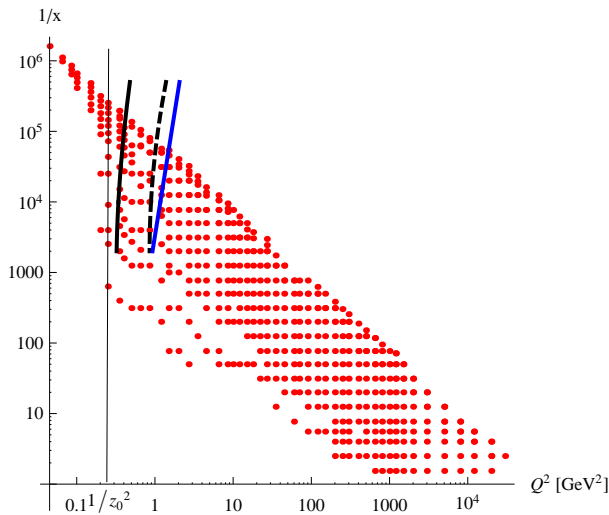}
  \caption{Saturation lines in the confining background for ${\bf b}_\perp =0 \ GeV^{-1}$ (black, dashed curve) and ${\bf b}_\perp =2 \ GeV^{-1}$ (black, solid curve) in comparison to the Golec-Biernat Wuesthoff result, (\ref{GBW}), (blue, solid) . The confining wall is at $1/z_0^2=1/4$.The dots are the measured HERA data. See text.}
  \label{satline}
  \end{center}
\end{figure}

Finally, we note that (\ref{transzendental}) defines the saturation line as a separatrix between the dilute and dense {\it wee-dipole}  environment. For fixed virtuality $Q^2$ (inverse dipole size squared) and impact parameter ${\bf b}_\perp$, Figure~\ref{satline} shows the rapidities at which the cross section saturates in the black solid and dashed curves, ie. when the condition in (\ref{transzendental}) is fulfilled. For comparison, the Golec-Biernat Wuesthoff result in (\ref{GBW}) is shown as the blue solid curve.
The points are the measured HERA data for the $F_2$ structure function. We note that (\ref{transzendental}) admits in general 2 distinct solutions for fixed $x,Q^2$ or $\chi$, but only the one with the largest $\chi$ is shown which is warranted by our approximations. The HERA points at the left of the confining wall $1/z_0^2=1/4$  are well within the confining region. The confined
holographic saturation lines are {\it stiff} in longitudinal energy as already noted in Figure~\ref{Qsatx} above.
For $b_\perp=0$ the HERA points to the far right of the saturation line are well within the perturbative or dilute {\it wee-dipole}  phase. Those close or to the left of the saturation line
corresponds to the saturated {\it wee-dipole}  phase. The closer they are to the confining wall
$1/z_0^2=1/4$ the less perturbative they are in nature. 
The holographic saturation lines show that a large swath of the measured points at HERA  which 
is well within the holographic saturation domain is sensitive to the impact parameter dependence 
${\bf b}_\perp$ of the saturation scale.
\\
\\
\\
\\
\section{Conclusions}\label{conclusion}

Dipole-dipole scattering in holographic QCD is purely imaginary at large rapidity $\chi$ which is a key feature of QCD. 
It follows from the t-channel exchange of closed-strings induced by a prompt longitudinal "electric" field.  The pomeron 
with N-ality 1 is a closed string exchange triggered by a stringy Schwinger mechanism. The creation process fixes 
the Pomeron slope, intercept and weight (``residue") in the elastic amplitude. From the open-closed string
duality, Gribov's diffusion follows from the  presence of a large "electric" acceleration or
Unruh temperature that causes the tachyonic mode of the dual open string to dimensionally reduce from 
$D$ to $D_\perp$ and diffuse.

In curved AdS space, the holographic direction is identified with the size of the dipole. The idea of Gribov diffusion appears as a string tachyon diffusion in both virtuality and transverse space.  In the conformal limit and for $D_\perp=3$, the dipole-dipole scattering 
amplitude and its related {\it wee-dipole}  density are found to be identical to the QCD results for onium-onium scattering using the
QCD BFKL pomeron. The results are readily extended to confining AdS with a wall,  and yield an explicit relation for the dipole
saturation momentum as a function of rapidity $\chi$ (or equivalently ${\rm ln}(1/x)$) and impact parameter ${\bf b}_\perp$.  For
large impact parameter, the holographic saturation momentum is closely related to the GBW saturation momentum~\cite{GolecBiernat:1998js}.

The dipole-dipole scattering amplitude in both conformal and confining AdS$_3$ is used to analyze the $F_2$ structure
function. A comparison with DIS data from HERA shows that the $x$ and $Q^2$ dependence of our holographic result are compatible with the data in the 1-pomeron exchange approximation, with no a priori need for an eikonal multi-pomeron resummation.  This conclusion is only qualitative as a more thorough study of the parameter space of the holographic model as well as the fitting accuracy are needed.

 \section{Acknowledgements}

We would like to thank Gokce Basar and Ho-Ung Yee for discussions.
This work was supported by the U.S. Department of Energy under Contracts No.
DE-FG-88ER40388.


\end{document}